\documentclass[lettersize,journal]{IEEEtran}
\usepackage{hyperref}
\usepackage{color}
\usepackage{epsfig}
\usepackage{array}
\usepackage{graphicx}
\usepackage{algorithm}
\usepackage{caption}
\usepackage{subcaption}
\usepackage{xspace}
\usepackage{minted}
\usepackage{tikz}
\usepackage{amsmath}
\usepackage{amsfonts}
\usepackage{mathtools}
\usepackage{siunitx}
\usepackage{tabularx} 
\usepackage{booktabs}
\usepackage{pifont}
\usepackage{wrapfig}
\usepackage{algpseudocode}
\newcommand{\cmark}{{\color{green!60!black}\ding{51}}} 
\newcommand{\xmark}{{\color{red!75!black}\ding{55}}} 

\begin{document}


\pagenumbering{arabic}
\newcommand{\Design}[0]{Rosemary\xspace}

\title{\Design: Efficient and Adaptable In-Storage Acceleration for Fully Homomorphic Encryption}

\author{Xuan Wang, Tianqi Zhang, Gabrielle De Micheli, Yujin Nam, Keming Fan, Augusto Vega, Minxuan Zhou~\IEEEmembership{Member,~IEEE}, Tajana Rosing~\IEEEmembership{Fellow,~IEEE}
\thanks{Xuan Wang, Tianqi Zhang, Gabrielle De Micheli, Yujin Nam, Keming Fan and Tajana Rosing are with the University of California San Diego.}
\thanks{Augusto Vega is with IBM Research, USA.}
\thanks{Minxuan Zhou is with Illinois Institute of Technology.}}

\markboth{IEEE TRANSACTIONS ON COMPUTER-AIDED DESIGN OF INTEGRATED CIRCUITS AND SYSTEMS}%
{Shell \MakeLowercase{\textit{et al.}}: A Sample Article Using IEEEtran.cls for IEEE Journals}

\newif\ifshowcomments
\showcommentstrue

\ifshowcomments
\newcommand{\todo}[1]{\textcolor{pink}{TODO: #1}}
\newcommand{\question}[1]{\textcolor{blue}{\sf Question: #1}}
\newcommand{\newquestion}[1]{\textcolor{red}{\sf Question2: #1}}
\newcommand{\minxuan}[1]{\textcolor{teal}{[\sf\bfseries    Minxuan: #1       ]}}
\newcommand{\Keming}[1]{\textcolor{purple}{Keming: #1}}
\newcommand{\changes}[1]{\textcolor{black}{#1}} 

\else 
    \newcommand{\todo}[1]{}
    \newcommand{\question}[1]{}
    \newcommand{\newquestion}[1]{}
    \newcommand{\help}[1]{}
    \newcommand{\minxuan}[1]{}
    \newcommand{\Keming}[1]{}
    \newcommand{\changes}[2]{}
\fi

\newcommand{\tikzcircle}[1]{\tikz[baseline=(char.base)]{
            \node[shape=circle,draw,fill=black,inner sep=1pt] (char) {\textcolor{white}{#1}};}}
\newcommand{\tikzcirclewhite}[1]{\tikz[baseline=(char.base)]{
            \node[shape=circle,draw,inner sep=1pt] (char) {#1};}}

\newcommand{\revise}[1]{\textcolor{black}{#1}}

\maketitle

\begin{abstract}
Fully Homomorphic Encryption (FHE) enables computation on encrypted data and is promising for privacy-preserving computing. However, FHE deployment remains constrained due to massive computational, memory, and data-movement overheads. Existing acceleration strategies often overlook the storage I/O bottleneck caused by the enormous size of FHE data in large-scale applications, as most prior works focus on small benchmarks that fit in main memory. Our studies show that state-of-the-art FHE systems spend 71\% to 85\% of their execution time moving data between storage and compute units rather than performing useful computation. To address the I/O bottleneck, we present \Design, a heterogeneous FHE acceleration system that incorporates a novel in-storage processing (ISP) architecture with compiler-level and algorithmic optimizations. Our key innovations include 1) an adaptable hardware-aware FHE engine inside a NAND flash-based solid-state drive (SSD) for efficient and cost-effective in-storage FHE processing, 2) a Throughput-Data-Aware Program Allocator (TDAP) that partitions FHE workloads into heterogeneous programs, and 3) a Compiler-Level Ciphertext-Aware Optimizer (CLCAO) that exploits ISP-friendly FHE kernels for low-cost ISP execution. Across a diverse set of FHE workloads, including encrypted deep learning inference and cloud database queries, \Design achieves geometric-mean speedups of $185.7\times$ over state-of-the-art CPU/GPU implementations and $3.7\times$ over prior FHE accelerators, while improving energy efficiency by up to $3.9\times$ and Energy-Delay-Area-Product (EDAP) by up to $16.5\times$.
\end{abstract} 


\section{Introduction}
\label{sec:intro}
As data storage and computation increasingly move to the cloud, preserving data privacy has become critical~\cite{rhychee,pathe}. Regulations~\cite{hipaa, gdpr} further mandate strong privacy protection. Fully Homomorphic Encryption (FHE)~\cite{bfv,bgv,ckks_scheme1} addresses this need by enabling computation on encrypted data without decryption. FHE is especially valuable for privacy-sensitive cloud applications such as deep learning inference, database analytics, and information retrieval.

However, existing FHE schemes suffer from prohibitive performance overhead. With highly customized FHE libraries~\cite{openfhe, seal, lattigo}, computations are still $10^4\times$ to $10^5\times$ slower than their unencrypted counterparts on CPUs~\cite{craterlake}. Fortunately, recent advances in ASIC accelerators~\cite{cinnamon, hercales} have dramatically improved the performance of FHE, which has narrowed the performance gap between FHE operations and their unencrypted counterparts to 2-4 orders of magnitude (e.g.,~\cite{sharp, craterlake}). Previous studies also demonstrated 2-3 orders of magnitude speedup on GPUs~\cite{cheddar, neo, tensorfheplus, fideslib} compared to CPU implementations.

Despite those advances, adopting FHE acceleration in real systems remains limited by system-level bottlenecks, where the enormous cost of data movement and heavy computational demands severely restrict achievable performance and scalability. Depending on the parameter configuration, ciphertexts may expand to $10^3\times$ to $10^5\times$ compared to the size of their corresponding plaintext version~\cite{ufc, smartpir, ciphermatch}. Consequently, the memory footprints of FHE workloads frequently reach hundreds of gigabytes or even terabytes, far exceeding typical main memory capacity. For example, FHE-enabled deep learning inference can require model weights up to hundreds of terabytes~\cite{ppml_22, ppml_22_2, multigpu}, while data-intensive workloads, such as database search~\cite{llm_database, fhe_database} and information retrieval~\cite{mulpir, ares}, may operate on datasets of even larger size.

\begin{figure}[t]
    \centering
    \epsfig{file=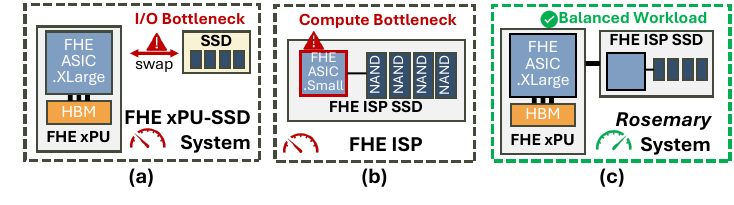, width=\columnwidth}
    \caption{Comparison of different FHE architectures. (a) FHE-host-accelerator only; (b) prior ISP designs for FHE; (c) \Design.}
    \label{fig::intro}
\end{figure}

Given the excessive memory footprint, naively connecting the host-side FHE accelerators with large memory to SSD (Figure~\ref{fig::intro}(a)), leads to a significant system bus bottleneck for I/O bound FHE computations. Prior work~\cite{smartpir, inspire} and our analysis in Section~\ref{sec:io} demonstrate $57\%$ to $85\%$ I/O overhead for overall execution time on CPU/GPU/ASIC systems. Motivated by this observation, in-storage-processing (ISP), which moves computation closer to storage, presents a promising approach. 
Several prior works have explored ISPs to accelerate FHE workloads~\cite{smartpir, inspire, ciphermatch, ares}. However, these works primarily adopt an ISP-only approach (Figure~\ref{fig::intro}(b)), where the limited compute and memory resources within SSDs hinder the adoption of such accelerators to support FHE with large parameter sizes and complex operation patterns.
As a result, existing designs either accelerate only lightweight FHE programs~\cite{smartpir, inspire, ciphermatch, ares}, or deliver sub-optimal performance~\cite{poseidon_ndp}. CIPHERMATCH~\cite{ciphermatch} targets exact string matching, while INSPIRE~\cite{inspire} and SmartPIR~\cite{smartpir} leverage in-storage PIR (Private Information Retrieval). Although Poseidon-NDP~\cite{poseidon_ndp} supports general FHE, its FPGA-based design requires over 1000s even for shallow MNIST inference, far behind ASIC accelerators. Consequently, ISP-only designs are largely limited to lightweight or specialized FHE workloads, while ASIC-only accelerators continue to suffer from high storage data-movement overheads, hindering efficient support for large-scale end-to-end FHE applications.

\textbf{Our goal} in this work is to architect an efficient and adaptable FHE acceleration system to tackle the computation and I/O challenges of large-scale privacy-preserving applications by exploiting heterogeneous computing with a novel in-storage processing accelerator integrated in host-side ASIC-accelerated FHE systems (Figure~\ref{fig::intro}(c)).
To achieve our goal of efficiently integrating ISP into a full-stack FHE system, we propose \textbf{\textit{\Design}}, which addresses challenges stemming from the hardware properties of FHE and the architectural characteristics of the storage devices via hardware-software co-design. First, we propose a high-performance ISP architecture with novel hardware components for accelerating general FHE operations in storage, including (i) reconfigurable FHE processing units for cost-effective processing of various kernels and (ii) a hierarchical on-chip buffer architecture for ciphertext and key management. Second, for the heterogeneous system, \Design incorporates compiler-driven optimizations to identify HE computations suitable for in-storage execution, partition workloads across the host and SSD, and optimize ciphertext and key movement for efficient heterogeneous execution. The key contributions of this work are as follows:
\begin{itemize}

\item We propose Rosemary, an efficient, cost-effective, and adaptable heterogeneous FHE acceleration system that exploits ISP for I/O challenges of large-scale applications. 


\item \Design introduces coordinated compilation innovations, including the \emph{Throughput-Data-Aware Program Allocator (TDAP)} for heterogeneous partitioning across the host and SSD, and the \emph{Compiler-Level Ciphertext-Aware Optimizer (CLCAO)} for memory optimizations.

\item We conduct workload characterization over FHE workloads and comprehensive design space explorations to derive new insights into I/O-compute tradeoffs. These insights motivate the key architectural choices in \textit{RosemarySSD}, enabling ISP-friendly execution and improved exploitation of SSD-internal parallelism.

\item Our evaluation shows that \Design delivers geometric-mean speedups of $185.7\times$ over state-of-the-art CPU/GPU implementations and $3.7\times$ over prior FHE accelerators~\cite{ufc}. It also improves energy efficiency by up to $3.9\times$ and reduces EDAP by up to $16.5\times$.



\end{itemize}

\section{Background}
\subsection{Fully Homomorphic Encryption}
\label{sec:fhe}
\subsubsection{FHE Basics}
\label{sec:fhe_basics}
Modern FHE schemes include CKKS~\cite{ckks_scheme1,ckks_scheme3},
BGV~\cite{bgv}, BFV~\cite{bfv}, and GSW~\cite{helib}. We mainly focus on CKKS for its wide popularity. CKKS is built on the ring learning-with-errors (RLWE) problem over the polynomial ring
$\mathcal{R}_Q=\mathbb{Z}_Q[X]/(X^N+1)$, where $N$ is the ring dimension and
$Q$ is the ciphertext modulus. A vector of raw data is first \emph{encoded} into a polynomial $\mathbf{m}\in\mathcal{R}_Q$, then encrypted under secret key $\mathbf{s}$ with small error $\mathbf{e}$ by sampling uniform random polynomial $\mathbf{a}$ and setting $\mathbf{b}=-\mathbf{a}\mathbf{s}+\mathbf{m}+\mathbf{e}$, producing ciphertext $\mathbf{ct}$:

\begin{equation}
\label{eq:rlwe_encrypt}
\mathbf{(c_0,\ldots,c_{N/2-1})}
\overset{\text{Encode}}{\longmapsto}
\mathbf{m}(X)=\sum_{i=0}^{N-1} k_i X^i
\;\overset{\text{Encrypt}}{\longmapsto}
\mathbf{ct}=(\mathbf{a},\mathbf{b})
\end{equation}

A function $f(\cdot)$ can be \emph{homomorphically evaluated} directly on the ciphertext without decryption. Such evaluation is expressed as a sequence of HE operations, including ciphertext-ciphertext and ciphertext-plaintext addition/multiplication (\texttt{HAdd}/\texttt{HMult}/\texttt{PMult}) and ciphertext slot rotations (\texttt{HRot}). To efficiently support large integers, CKKS uses residue number system (RNS) to represent $Q=\prod_{i=0}^{L-1} q_i$ as a product of $L$ word-sized primes, decomposing a degree-$N$ polynomial into an $L\times N$ layout, where each row is a \emph{limb} associated with prime $q_i$ and can be viewed as a polynomial.

\subsubsection{Main Functions of CKKS}
\label{sec:fhe_operations}

Homomorphic operations are modular arithmetic on ciphertext polynomials. \texttt{HAdd} and \texttt{PMult} use modular element-wise operations, \texttt{HRot} applies an automorphism that maps coefficient $i$ to $ik \bmod N$, corresponding to cyclic slot rotations, and \texttt{HMult} uses modular multiplication and addition. To accelerate polynomial multiplication in $\mathcal{R}_Q$, CKKS uses the NTT, which converts RNS limbs to the evaluation domain so negacyclic convolution becomes coefficient-wise multiplication, reducing complexity from $O(N^2)$ to $O(N\log N)$~\cite{fft1,fft2}. The iNTT converts results back only when coefficient-domain access is needed.

\texttt{HMult} and \texttt{HRot} change the secret key of the result ciphertext, so CKKS invokes \emph{key switching} to restore the original key. This uses a large evaluation key, often hundreds of MB~\cite{mad}, $\mathsf{evk}\in\mathcal{R}_{PQ}^{2\times\text{dnum}}$, where $P$ is an auxiliary modulus and $\textit{dnum}=\left\lceil \frac{L}{K}\right\rceil \ge 1$ is the decomposition number. In RNS form, the evaluation key consists of $\textit{dnum}$ decomposition components, each containing a pair of $(L+K)\times N$ matrices. Key switching also requires base conversion (\emph{BConv}) to match moduli. \emph{BConv} mainly performs a matrix-matrix multiplication between an input polynomial and a fixed $K\times L$ base table, but it requires coefficient-wise representation. As a result, $\mathrm{iNTT}\rightarrow\mathrm{BConv}\rightarrow\mathrm{NTT}$ is a common sequence in key switching and requires coefficient-wise access to the full polynomial. To sustain an unbounded multiplicative budget, FHE uses \emph{bootstrapping} to refresh the accumulated noise budget from computations and restore depth. However, bootstrapping is very expensive, dominated by a sequence of NTT/iNTT transforms and repeated keyswitching~\cite{boot1,boot2}.

\begin{figure*}
    \centering
    
    \begin{minipage}[t]{0.33\textwidth} 
        \centering
        \includegraphics[width=0.98\linewidth]{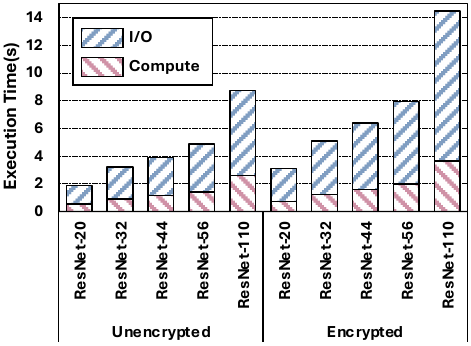}
        \caption{The execution time of ResNet with Encrypted and Unencrypted weights.}
        \label{fig::ssd_overhead}
    \end{minipage}\hfill
    \begin{minipage}[t]{0.64\textwidth} 
        \centering
        \includegraphics[width=0.93\linewidth]{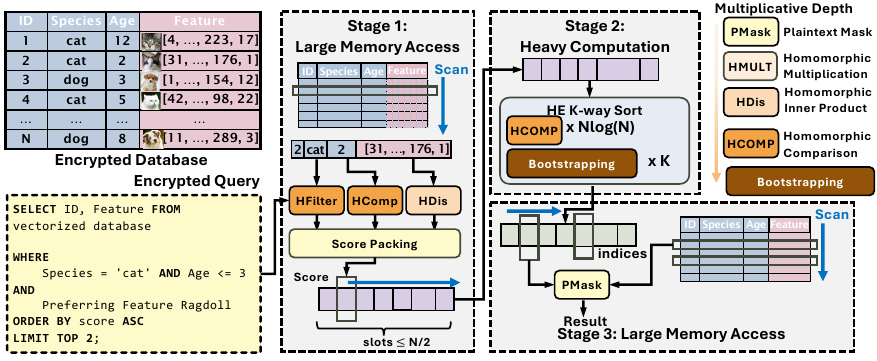}
        \caption{Code example for heterogeneous processing.}
        \label{fig::hetergenous_code}
    \end{minipage}
\end{figure*}

\subsection{In-Storage Processing}
\subsubsection{SSD Basics}
A NAND flash-based SSD comprises a controller, host interface, on-device DRAM, and multiple flash channels connected to NAND dies. The controller manages the overall system, while a Direct Memory Access ($\text{DMA}$) engine supports high-throughput host-SSD transfers. An embedded processor runs firmware for Flash Translation Layer (FTL), including logical-to-physical (L2P) address translation and garbage collection~\cite{gc1}. To access flash efficiently, the SSD employs multiple per-channel flash controllers so that channels can operate in parallel and provide high aggregate bandwidth. 

\subsubsection{ISP Architecture} 
\label{sec:isp}
In-Storage Processing (ISP) reduces memory-hierarchy bottlenecks by moving data-intensive computation closer to storage. By integrating compute units within the SSD controller or near-chip/die, ISP exploits the SSD’s internal bandwidth and parallelism to reduce host-memory pressure and data movement. Prior non-FHE ISP designs have used programmable logic~\cite{csd1, smartssd, smartpir} or specialized hardware accelerators~\cite{megis, beacongnn, ndsearch} to accelerate I/O-bound workloads such as bioinformatics, retrieval-augmented generation, graph neural networks, and databases.

\section{Motivation}
\label{sec:motivation}
\subsection{I/O Overhead Analysis}
\label{sec:io}
FHE drastically increases memory demand due to slot-based encoding and encryption, while FHE computation continues to accelerate, targeting on bridging the gap with unencrypted counterparts. The $10^3$--$10^5\times$ memory overhead therefore makes data movement a dominant bottleneck. This growing imbalance motivates an FHE architecture to mitigate I/O constraints. Figure~\ref{fig::ssd_overhead}(a) breaks down ResNet inference latency into computation and SSD I/O based on our FHE accelerator performance model with a cold-start HBM. In a host-accelerator-only baseline, SSD accesses account for about $70\%$ of execution time for inference with unencrypted weights and $75\%$ for inference with encrypted weights. Prior studies also report substantial I/O overheads, reaching $57\%$ and $85\%$ in CPU- and GPU- based systems~\cite{inspire,smartpir}. FHE inference quickly exceeds main-memory capacity. Existing FHE inference systems require up to 512GB RAM due to hundreds of GBs of data, while BERT-Base and Llama3-8B can demand about 1.5TB and 112TB, respectively~\cite{ppml_22, ppml_22_2, multigpu, llm_nexus}. Even optimized LLM weights alone can occupy 982GB~\cite{multigpu}, making storage I/O the scalability bottleneck.


\subsection{Limitations of Prior Approaches}
\label{sec:limitations}
Table~\ref{tab:prior_compare} highlights the trade-offs for prior FHE accelerators. Host-accelerators support general workloads and large parameters but incur high data-movement overhead when ciphertexts exceed main memory~\cite{sharp, ufc, craterlake}. Scale-out systems provide greater compute and memory capacity, yet still suffer substantial inter-device I/O~\cite{hydra, cinnamon}. ISP accelerators reduce data movement but mainly target lightweight workloads and smaller parameters (e.g., $N=2^{13}$ to $2^{14}$~\cite{inspire, smartpir, ares}. \Design bridges these approaches through heterogeneous execution that balances compute capability and I/O efficiency.

\begin{table}[b]
\centering
\caption{Trade-offs of prior FHE acceleration approaches.}
\label{tab:prior_compare}
\fontsize{8}{9}\selectfont
\setlength{\tabcolsep}{2.5pt}
\renewcommand{\arraystretch}{1.05}
\begin{tabularx}{\columnwidth}{
    >{\raggedright\arraybackslash}p{0.2\columnwidth}
    c c c
    >{\raggedright\arraybackslash}X
}
\toprule
\textbf{Approach} & \shortstack{\textbf{Low}\textbf{I/O}} & \shortstack{\textbf{Large}\textbf{$N$}} & \textbf{General} & \textbf{Trade-offs} \\
\midrule
Host-Accel~\cite{sharp,ufc}
& \xmark & \cmark & \cmark
& High compute throughput, but high data movement. \\
\midrule
ISP-only~\cite{inspire,smartpir,ares}
& \cmark & \xmark & \xmark
& Low data movement, but limited compute capability. \\
\midrule
Scale-out~\cite{hydra,cinnamon}
& \xmark & \cmark & \cmark
& Increased parallelism and aggregated memory, but persistent I/O overhead. \\
\midrule
\textbf{\Design} (this work)
& \cmark & \cmark & \cmark
& Balanced compute and I/O through heterogeneous execution. \\
\bottomrule
\end{tabularx}
\end{table}

\subsection{Heterogeneous Computing with ISP for FHE}
\label{sec:heter_motivation}

To overcome the scalability and generality limitations of prior FHE ISP-only accelerators, \Design adopts a heterogeneous architecture that jointly exploits ISP resources and host-side accelerators. ISP is beneficial when its computation fits within the SSD’s critical I/O path. Specifically, with $b$ parallel flash channels, NAND-read latency $t$ cycles, page size $p$ bytes, per-operation latency $n$ cycles, and ciphertext size $cipher_{size}$, the maximum number of in-storage operations that can be hidden by I/O is approximately $\frac{cipher_{size} \times t}{b \times p \times n}$.

Figure~\ref{fig::hetergenous_code} gives an example of how heterogeneous processing can benefit a FHE vectorized database search workload. This example queries for a cat candidate younger than $3$ years old, preferably a Ragdoll, from a vectorized database with $n$ entries. HE comparison \cite{engorgio, llm_database}, HE gate \cite{engorgio}, and HE inner product/HE distance~\cite{llm_nexus,nn_gazelle} are offloaded to the ISP because they involve intensive direct access to the encrypted database and have shallow multiplicative depth. HE sorting~\cite{fhe_database,engorgio} remains on the host due to its deep multiplicative depth, frequent rotations, and potential need for multiple bootstrappings. The identified indices and distance scores are returned to the host for sorting, and the sorted results are then sent back to the SSD for final retrieval from the encrypted database. In summary, ISP reduces data movement by computing near storage, but supporting general-purpose FHE remains challenging under SSD resource constraints.

\section{\Design Design}
\label{sec:design}
\Design is an efficient heterogeneous ISP system that supports broad FHE parameters and end-to-end applications.
owever, building such a system presents both hardware and software challenges. On the hardware side, the host provides abundant compute resources but is constrained by storage I/O bandwidth and memory capacity. SSDs, in contrast, operate closer to the data but offer limited logic and buffering resources and must accommodate flash-specific constraints such as page-granularity access. These limitations complicate support for diverse FHE operators and parameters, particularly because FHE polynomials can occupy several MBs. On the software side, the system must determine when ISP offloading is beneficial, coordinate execution between the host and SSD, and maintain generality across workloads. \Design addresses these challenges through hardware–software co-design, combining a \Design SSD architecture with algorithmic and compiler optimizations to integrate a cost-effective ISP accelerator into a heterogeneous execution flow.

\subsection{\Design System}
\label{sec:heter_system}
\begin{figure}[t]
    \raggedleft
    \epsfig{file=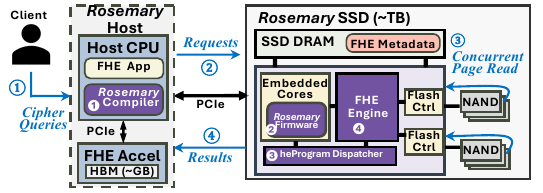, width=\columnwidth}
    \caption{\Design Overview.}
    \label{fig::overview}
\end{figure}

Figure~\ref{fig::overview} presents an overview of \Design system. \Design integrates an ISP-enabled SSD, \textit{RosemarySSD}, with a high-performance host-side accelerator, \textit{FHE Accel} (\tikzcircle{2}), to reduce host--SSD data movement while preserving execution flexibility and throughput without overburdening SSD resources. \textit{RosemaryHost} consists of \textit{Host CPU}, \textit{FHE Accel}, and \textit{RosemaryCompiler} (\tikzcircle{1}). At compile time, \textit{RosemaryCompiler} statically analyzes the FHE program and uses the Throughput-Data-Aware Program Allocator (TDAP, Section~\ref{sec:tdap}) to partition operations between \textit{FHE Accel} and \textit{RosemarySSD}. It then generates the instructions required for end-to-end execution and orchestrates the heterogeneous system. At runtime, the host follows the precomputed allocation to manage data transfers and issue execution commands over PCIe.

When \textit{RosemaryHost} receives the user query ciphertexts, \textit{Host CPU} issues ciphertext requests (\tikzcirclewhite{1}) and the precompiled mProgram (Section~\ref{sec:e2e}) to invoke SSD-Segment Execution of the ciphertext-heavy and data-movement-intensive program segments on the \Design Engine (\tikzcircle{3}) within the controller. During this SSD-Segment Execution, the SSD concurrently reads required FHE polynomials from the NAND flash backend across all channels (\tikzcirclewhite{3}) and returns the processed ciphertexts (intermediate) results (\tikzcirclewhite{4}) to \textit{FHE Accel} in \textit{RosemaryHost} for the compute-and-memory-bound segments that demand high on-chip throughput. This design embeds specialized FHE support within a commodity SSD controller, whose detailed architecture is described in Section~\ref{sec:isp}. Through hardware-software co-design and heterogeneous execution, \Design efficiently supports diverse FHE workloads across a wide range of parameter settings. 

\section{RosemarySSD}
\label{sec:isp} 
\texttt{RosemarySSD} is the SSD-side architecture that enables efficient FHE execution within the heterogeneous \Design system. \texttt{RosemarySSD} extends a commodity SSD controller with four key components: \tikzcircle{1} an FHE Engine, \tikzcircle{2} a scratchpad eDRAM, \tikzcircle{3} a heProgram Dispatcher, and \tikzcircle{4} the Rosemary FTL. It is designed to address three core challenges: (1) tight resource constraints in a low-power, high-capacity SSD, (2) adaptability across diverse FHE parameter settings and workloads, and (3) limited off-chip DDR DRAM bandwidth.

\subsection{FHE Engine Architecture}
\label{sec:fhe_engine}
\begin{figure}[b]
    \raggedleft
    \epsfig{file=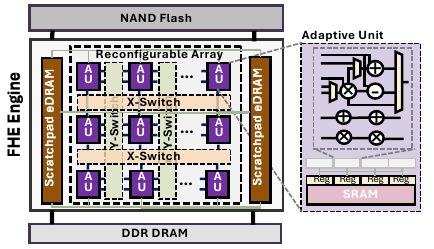, width=\columnwidth}
    \caption{RosemarySSD overview and Adaptive Unit.}
    \label{fig:design_overview_engine}
\end{figure}

\Design engine centers on a 2D array of Adaptive Units (AUs), shown in Figure~\ref{fig:design_overview_engine}, that unifies the dominant FHE kernels on a single reconfigurable datapath. Specifically, each AU contains multiple lanes of butterfly units~\cite{sapphire} and modular adders and multipliers. During execution, FHE polynomials are distributed across the lanes of all processing elements (PEs). When the polynomial size exceeds the total lane count, coefficients are streamed through the same lanes over multiple rounds. To reduce SRAM port pressure, \Design forwards intermediate data across processing units through vector chaining~\cite{craterlake, cray, vector_chain}, improving locality while sustaining throughput. Because \Design implements arithmetic primitives shared across major FHE schemes, its architecture can be extended to support schemes beyond CKKS with limited scheme-specific modifications.

A key advantage of the AU-based design is that it avoids kernel-specific logic and datapaths while remaining adaptable across FHE parameter and workload settings. Prior accelerators often employ dedicated base-conversion (BConv) units for key switching~\cite{sharp, ufc}. In contrast, \Design reuses the (i)NTT datapath, whose modular MAC, addition, and multiplication operators already support BConv. This sharing is efficient because BConv typically executes sequentially within the \textit{iNTT-BConv-NTT} pipeline rather than concurrently with (i)NTT. Through lightweight multiplexing, the same arithmetic units can also operate as element-wise modular adders or multipliers, efficiently supporting shallower workloads with different operation mixes. This flexibility is particularly valuable for information-retrieval workloads such as PIR~\cite{mulpir, ares}, which often have a higher proportion of element-wise operations than key-switching-intensive workloads. 

\Design centralizes the AU array in a single SSD-level FHE engine instead of distributing smaller accelerators across flash channels. This organization better matches SSD access granularity and FHE execution characteristics. NAND accesses are page-granular, whereas FHE polynomials are substantially larger (e.g., 256,KB) and span many pages and channels. A centralized engine can gather pages from all flash channels without cross-engine coordination, which is particularly important for kernels such as key switching, where the \textit{Decompose} $\rightarrow$ \textit{iNTT} $\rightarrow$ \textit{BConv} $\rightarrow$ \textit{NTT} sequence operates over the full ciphertext. In contrast, per-channel engines fragment the datapath, create load imbalance, and require costly replication of buffers and arithmetic units. Centralization also enables buffering. A centralized engine can buffer the $\alpha$ limbs from \textit{Decompose} and generate the remaining \textit{ModUp} limbs entirely on chip~\cite{mad}. Section~\ref{sec:ftl} describes the limb-aware data layout.

To keep AU design under the area and power budgets of the SSD-controller, \Design adopts Geometry-Constant NTT (GC-NTT)~\cite{cgntt1} to reduce permutation complexity. The remaining permutations are decomposed across the 2D PE array into x-axis switching (\textit{xshuffle}), y-axis shuffling (\textit{yshuffle}), and intra-PE shuffling (\textit{rshuffle})~\cite{ntt_shuffle}, implemented by the X/Y-Switch interconnects. Since all PEs can perform these transfers independently at each shuffle step, \Design pipelines them across NTT operations to sustain throughput. We support automorphism over the same 2D interconnect~\cite{ufc}, use an HE-friendly modular multiplier~\cite{fhendi}, generate twiddle factors on the fly, and store only half of each KSK while regenerating the pseudo-random half with a PRNG~\cite{craterlake, helib, sapphire}. 

\subsection{RosemarySSD Memory}
\label{sec:fhe_engine_memory}

\begin{wrapfigure}{l}{0.45\columnwidth}
    \centering
    \epsfig{file=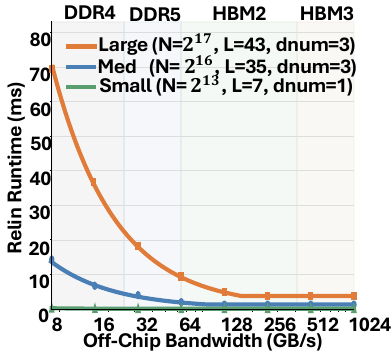, width=0.45\columnwidth}
    \caption{Performance comparison between different on-chip bandwidths.}
    \label{fig::ksk}
\end{wrapfigure}

\textit{RosemarySSD} uses a four-level memory hierarchy to keep hot data near the AU array and minimize off-chip accesses, as shown in Figure~\ref{fig:design_overview_engine}. Each AU uses a local 1R1W RegFile for operands and partial sums, while a shared \revise{32MB} SRAM buffers intermediate polynomials. The 256--512MB SRAM capacities used in prior accelerators~\cite{sharp, ufc} are impractical for an SSD controller.

However, during key switching, a small SRAM buffer forces tens of MB of KSK data to stream from off-chip memory, making performance highly sensitive to the limited DDR bandwidth of SSDs (Figure~\ref{fig::ksk}). \Design therefore adds a 64MB, 8-bank eDRAM scratchpad directly accessible by the AU array. SRAM buffers active intermediates, while eDRAM retains hot keys and overflow data, reducing off-chip accesses and NAND write-backs. In this hierarchy, SRAM serves as a fine-grained buffer for active intermediates, while eDRAM provides a larger storage layer for hot keys and overflow data that cannot fit in SRAM. eDRAM offers substantially higher density and lower cost under SSD power and thermal constraints, making large-capacity on-chip storage practical inside the controller. eDRAM refresh overhead is negligible for long-running FHE kernels: its $10^3$-cycle retention time and approximately 34-cycle parallel refresh latency can be hidden by computation~\cite{edram3}. Its refresh energy also remains substantially lower than that of an equivalently sized SRAM.

\begin{figure}[b]
    \raggedleft
    \epsfig{file=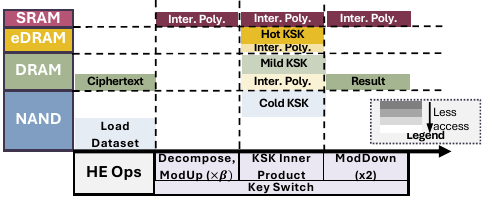, width=\columnwidth}
    \caption{RosemarySSD dataflow.}
    \label{fig:fhe_engine_dataflow}
\end{figure}

\subsection{Rosemary Dataflow}
\label{sec:fhe_engine_dataflow}
Figure~\ref{fig:fhe_engine_dataflow} illustrates the \textit{RosemarySSD} dataflow using key switching as an example. Before key switching, the input ciphertext from \textit{RosemaryHost} is stored in LPDDR and fed to the FHE engine, while the large FHE dataset is streamed from NAND flash into the eDRAM scratchpad through the flash channels. After earlier HE operations and scaling, the ciphertext enters the key-switching flow, which begins with decomposition and modulus raising (\textit{ModUp}). Decomposition expands the original $l$ polynomials into $\beta$ digits, each containing $\alpha$ polynomials, and the resulting intermediate polynomials are buffered in SRAM. The engine then executes the KSK inner-product step, where the much larger evaluation keys (KSKs) are supplied directly from the on-chip eDRAM scratchpad. If the required KSK is not resident in eDRAM, \textit{RosemarySSD} fetches it from LPDDR DRAM. If it is also absent there, it is retrieved from NAND flash.  When SRAM capacity is exceeded, intermediate polynomials overflow first into eDRAM, and further spill into LPDDR only if eDRAM capacity is exhausted. In this design, SRAM serves as a fine-grained intra-operation buffer, while eDRAM holds larger hot keys and streamed operands, allowing the AU array to sustain throughput without being throttled by off-chip bandwidth. This hierarchy keeps frequently reused keys and intermediates close to the 2D AU array, reducing repeated off-chip accesses across the \textit{iNTT-BConv-NTT} sequence.

\section{Compilation and Optimization}
\label{sec:scheduling}
\begin{figure}[b]
    \centering
    \epsfig{file=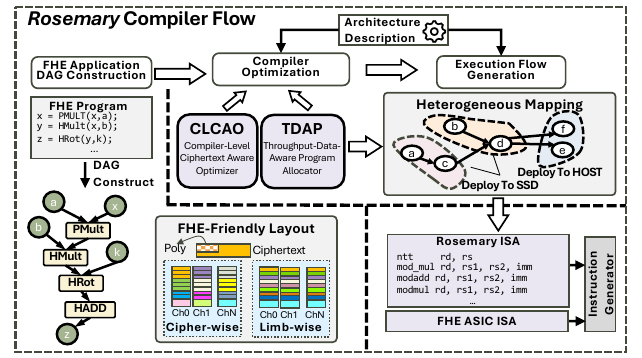, width=\columnwidth}
    \caption{Compilation flow of \Design.}
    \label{fig::compiler}
\end{figure}

To fully exploit \Design's heterogeneous architecture, we propose a hardware-aware compilation flow that maps FHE programs across the execution platform. Figure~\ref{fig::compiler} shows the overview of the proposed compilation stack, where the input to this flow is a weighted DAG representation of the FHE program. Specifically, each node corresponds to an FHE primitive operation (e.g., \textit{HMult}, \textit{HAdd}, \textit{Boot}), where the node weight captures the computational cost under the corresponding FHE parameters (e.g., ring dimension~$N$, modulus~$|q|$, and depth~$l$). Each edge represents ciphertext data dependencies, and its weight corresponds to the input or output ciphertext size in bytes. The architecture configuration of \Design is taken as an input for the compiler, which is used to statically estimate the execution latency and resource usage of each HE operation, enabling static scheduling given the regularity of FHE.

A Compiler-Level Ciphertext-Aware Optimizer (CLCAO) then applies optimization passes to convert operations in FHE DAG into ISP-friendly kernels and computation flows to improve the efficiency of execution using restricted ISP hardware resources. Then, a Throughput-Data-Aware Program Allocator (TDAP) optimizes the computation partition across the host-side \textit{FHE Accel} and \textit{RosemarySSD} using a cost model that captures compute throughput, memory capacity, and data-movement overheads. Finally, the backend generates the corresponding execution formats, including mPrograms for \textit{RosemarySSD} and APIs for the communication between host and SSD to enable seamless cooperative execution across the heterogeneous FHE system.

\subsection{Ciphertext-Aware Optimizer}
\label{sec:clcao}
\Design incorporates a Compiler-Level Ciphertext-Aware Optimizer (CLCAO) that reduces data movement from off-chip LPDDR to ISP processing logic through two memory-capacity- and bandwidth-aware optimizations for FHE kernels and computation flows. The first optimization is rotation-key decomposition. While \Design uses its SRAM-eDRAM hierarchy to keep hot data close to the AU array, the working set of rotation keys (i.e., KSK in Figure~\ref{fig:fhe_engine_dataflow}) can still exceed on-chip capacity, creating LPDDR bandwidth pressure that lowers throughput. CLCAO therefore analyzes the required rotation set, $\mathcal{R}$, and removes keys through rotation decomposition. Specifically, it decomposes a rotation $r$ into a short chain of $r=s_1+s_2+\cdots+s_k$ rotations using hot keys $s_1,\ldots,s_k$, which are already in eDRAM. This reduces the number of distinct rotation keys for the active working set. The decomposition is applied only when the key read cost exceeds the added compute cost, when $\tau_{\text{key}} > m\cdot\tau_{\text{rot}}$, where $\tau_{\text{key}}$ is the latency of reading the additional rotation key, $\tau_{\text{rot}}$ is the latency of one additional \textit{rot\_relin}, and $m$ is the number of extra rotation steps introduced by the decomposition. 


Another optimization is static ciphertext placement and scheduling, which exploits the regular structure of FHE programs to determine data placement and execution order ahead of time. Because encrypted data cannot directly drive data-dependent control flow, FHE workloads are typically static with fixed execution structure. CLCAO therefore performs compile-time liveness analysis to track ciphertext production, reuse, and lifetime, assigning each live ciphertext a cost of evicting and reloading based factors such as size and level. With this cost model, CLCAO optimizes data placement and execution order to minimize expensive evictions from the SRAM--eDRAM hierarchy. For example, in vectorized database score summation, a straightforward schedule may first calculate many score ciphertexts before defer accumulation, increasing memory pressure and causing short-lived intermediates to spill to LPDDR. CLCAO instead performs partial reductions earlier, consuming scores closer to their production time. This shortens their live ranges, reduces costly spills, and improves sustained throughput.

\subsection{Program Allocator}
\label{sec:tdap}

\Design requires effective allocation of program segments between the host \textit{FHE Accel} and \textit{RosemarySSD} to jointly balance computation cost and I/O cost. However, the computation cost of each node is placement-dependent and thus unknown until the partition is fixed. We therefore leverage the Binary-Min-Cut~\cite{mincut} algorithm and propose the \emph{Throughput-Data-Aware Program Allocator (TDAP)}, which performs cost-aware heterogeneous partitioning over the weighted DAG to assigns each segment based on placement-dependent execution cost and host--SSD I/O overhead. TDAP improves end-to-end throughput by jointly considering segment execution efficiency and ciphertext I/O movement between host and SSD.

First, TDAP estimates the execution cost of each homomorphic operation on \textit{FHE Accel} and \textit{RosemarySSD}, as well as the associated ciphertext I/O cost:
\begin{equation}
D_i(d)=C_{i,\mathrm{comp}}(d)+C_{i,\mathrm{mem}}(d)+C_{i,\mathrm{key}}(d).
\end{equation}
where $T_{ij}^{\mathrm{comm}}=\frac{b_{ij}}{B_{\mathrm{PCIe}}}$, and $D_i(d)$ is the raw execution cost of operation $i$ on device $d \in \{\mathrm{Host},\mathrm{SSD}\}$, and the three terms capture arithmetic execution, ciphertext buffer, and scratchpad memory traffic, and KSK access overhead, respectively. $T_{ij}^{\mathrm{comm}}$ is the I/O cost for the edge $(i,j)$, where $b_{ij}$ is the transferred ciphertext volume and $B_{\mathrm{PCIe}}$ is the PCIe bandwidth. A caveat is that, for SSD execution, the cost of key-switching-heavy operators (e.g., \textit{HMult} and \textit{HRot}) depends on whether their predecessor nodes are also placed on the SSD. For tractable static estimation, we assume predecessors are assigned to the same execution domain as the current node.

Using these estimates, TDAP selects the placement $x_i \in \{\mathrm{Host},\mathrm{SSD}\}$ with the Binary-Min-Cut~\cite{mincut} algorithm such that it minimizes
\begin{equation}
\min_{\{x_i\}}
(\sum_{i\in\mathcal{V}} D_i(x_i)
+
\sum_{(i,j)\in\mathcal{E}} \mathbf{1}[x_i \neq x_j]\cdot T_{ij}^{\mathrm{comm}}).
\label{eq:tdap_obj}
\end{equation}
The first term favors the device that executes each segment more efficiently, while the second penalizes cuts across large ciphertext boundaries. As a result, TDAP optimizes end-to-end throughput at the program level rather than making isolated per-segment decisions. To ensure tractable partitioning, we follow the regularized graph-construction constraints in~\cite{mincut}, which avoid an NP-hard formulation. Specifically, all weights are required to be positive, and we assume $D_i(\text{SSD}) \geq D_i(\text{Host})$, reflecting the fact that the SSD is less powerful than the host due to power budget constraints. TDAP performs node-level placement over the FHE DAG, allowing multiple host and SSD subgraphs. Consecutive operators mapped to the same domain are coalesced into an \textit{mProgram}, while cross-domain dependencies incur ciphertext-transfer costs, discouraging excessive host--SSD switching.



\subsection{FHE-Friendly Interleaving and FTL}
\label{sec:ftl}
The data layout within the SSD has a substantial impact on SSD channel bandwidth utilization. Since FHE ciphertexts are large RNS objects, even a single polynomial/RNS limb can span multiple SSD pages. \Design therefore interleaves the pages of each limb across multiple channels, enabling the FHE engine to leverage aggregated internal channel bandwidth during processing, as shown in Figure~\ref{fig::compiler}. \revise{Specifically, FHE ciphertexts and evaluation keys are initially written through the normal NVMe and FTL path. During pre-computation stage, each FHE object are written according to the channel-interleaved layout through API commands (Section~\ref{sec:sys_support} and publishes its object descriptor only after all object pages and metadata have been persisted. An update is written out of place as a new immutable object generation and is relayed out before use. After the new generation is complete, the controller atomically changes the active descriptor, and the old generation becomes reclaimable through normal garbage collection. Rosemary therefore does not support in-place modification of an object generation while it is being processed.}


The regularity of \Design's data layout also simplifies FTL metadata. Because each polynomial limb is distributed across channels evenly, the controller does not need page-granularity L2P entries. Instead, it only stores a compact object descriptor and derives page locations analytically from the base address and interleaving pattern. For example, with \(N=65536\) and 4-byte coefficients,  16~KiB SSD pages, a ciphertext with two polynomials and \(L=30\) limbs occupies 15MB, or 960 pages. Under a conventional page-level FTL with 4-byte entries, this requires $\approx$3.75KB of metadata. In contrast, \Design only needs a small descriptor, e.g., two 4-byte base addresses for the two polynomials with a few bytes for the limb count and layout rule, for a total of 16B per ciphertext. Thus, \Design reduces metadata from about 3.75~KB to only 16~B per ciphertext, about \(240\times\), while still enabling full internal channel-bandwidth utilization. Prior works also leverage customized FTL approaches to release DRAM capacity~\cite{megis, ndsearch, beacongnn}.

\section{\revise{System Support}}
\label{sec:sys_support}
\subsection{\revise{End-to-End System Design}}
\label{sec:e2e}
To enable end-to-end heterogeneous FHE execution, \Design defines a host-\textit{RosemarySSD} communication API, together with \textit{mProgram}, an instruction interface for the ISP FHE engine. \Design exposes a uniform API (Table~\ref{tab:api}) to enable input and intermediate ciphertext communication between the host and \textit{RosemarySSD}. This API is tightly coupled with our heterogeneous compilation flow. After the compiler selects segments offloaded to SSD and generates their corresponding \textit{mProgram}, the runtime uses these commands to transfer ciphertexts and trigger execution on the ISP FHE engine. Thus, the API bridges compiler-driven partitioning and end-to-end heterogeneous execution.

\begin{table}
\centering
\caption{\Design API Summary}
\small
\label{tab:api}
\setlength{\tabcolsep}{2pt}
\begin{tabularx}{\columnwidth}{@{}l p{0.28\columnwidth} X@{}}
\toprule
\textbf{Type} & \textbf{API} & \textbf{Description} \\
\midrule
Init & \textit{Prog\_Set(op, mprog)} 
& Store compiled mProgram to device. \\

Run & \textit{Send(ctxt, mprog)} 
& Send encrypted query to device. \\

Run & \textit{Recv(ctxt)} 
& Retrieve encrypted results from device. \\
\bottomrule
\end{tabularx}
\end{table}

When the FHE program is offloaded to the SSD, we leverage the \revise{vendor-reserved} ONFI~\cite{onfi1, onfi2} commands to support two control primitives: \textit{config()}, which sets FHE parameters and configures the AU array, and \textit{init()}, which starts execution. Upon \textit{init()}, the \textit{heProgram Dispatcher} (Section~\ref{sec:fhe_engine}) drives the FHE engine according to an \textit{mProgram}, a sequence of simple fixed-format commands summarized in Table~\ref{tab:mprog}. \revise{Reserved ONFI commands are internal controller micro-operations generated after command validation, FTL translation, and channel scheduling; they are not exposed directly to host software. This preserves the protection and resource-management responsibilities of the conventional SSD firmware.} All command fields are statically resolved by the compiler, referring only to logical NAND addresses or pre-allocated buffer/scratchpad locations. Thus, no dynamic allocation or runtime address computation is required beyond standard FTL translation. This makes \textit{mProgram} compact. For example, an ISP segment of Llama70B needs only 3.8MB of \textit{mProgram}, even less than the size of a single ciphertext.

\begin{table}
    \centering
\caption{\textit{mProgram} Command Definition}
\label{tab:mprog}
\small
\label{tab:placeholder}
    \begin{tabularx}{\columnwidth}{ll}\toprule
         \textbf{}{Instruction}& \textbf{Operands}\\\midrule
         \{NTT, iNTT\}& A\_addr, Out\_addr\\
         \{MODADD, MODMUL, MODMAC\}& A\_addr, B\_addr, Out\_addr\\
         AUTOMORPH& A\_addr, rotIdx, Out\_addr\\
         \{LD, ST\}& Addr, Type\\ \bottomrule
    \end{tabularx}   
\end{table}

\subsection{\revise{SSD Management and Control Tasks}}
\label{sec:ssd_manage}
\revise{\Design preserves the conventional SSD control plane: the NVMe frontend, embedded cores, FTL, ECC engines, GC, and NAND scheduler retain their existing roles. The FHE engine operates as a separate acceleration plane, consuming data after ECC decoding and issuing page requests through the existing FTL and flash scheduler, without involving embedded cores in FHE arithmetic or modifying NAND-channel control. Additional state is limited to the mProgram command queue, FHE configuration registers, a small object-descriptor cache, buffer-ownership state, and arbitration logic between accelerator and conventional I/O requests. Once FHE metadata is placed, \Design enters a dedicated ISP mode and no longer serves conventional host I/O, following prior work~\cite{ciphermatch, smartssd, megis}. Because \Design only reads NAND data and does not generate NAND writes, it introduces no additional write-induced endurance overhead~\cite{lifetime1} or write-side management such as garbage collection and wear leveling~\cite{gc1, megis, wl1}.
}

\section{Methodology}
\subsection{Hardware Implementation and System Modeling}
We develop an event‑driven, cycle‑accurate in-house simulator based on MQSim~\cite{mqsim}. The simulator models the latency of traditional SSD components, including embedded processor cores, SSD DRAM, and the multi‑level flash backend, as well as the customized multi‑level processing engines introduced in our design. The mPrograms, which encode all basic FHE operators and data-movement kernels as ISA-level commands, are used to drive our simulation. We implemented and synthesized the major logic units with a 32-bit word length at the RTL level using the ASAP7 7nm process design kit (PDK)\cite{asap7} and support bootstrapping with double-scaling operations~\cite{sharp}. The technology node is selected based on the 8nm and 5nm technology nodes for controllers in existing commercial SSD products~\cite{990pro4tb}. We provision a total of 64 MB eDRAM and 32 MB SRAM and use FinCACTI~\cite{fincacti} to model the interconnects, register files, SRAM, and eDRAM components with the published information \cite{edram1, edram3}. The FHE engine is composed of 4$\times$4 AU arrays, where each AU contains 128 lanes. We model the SSD with the configuration in Table~\ref{tab:sim_config} based on a commercial SSD design~\cite{990pro4tb}.

\begin{table}
\caption{SSD Simulation Configuration.}
\centering
\resizebox{\columnwidth}{!}{
\begin{tabular}{|c|l|}
\hline
\textbf{Component} & \textbf{Configuration} \\
\hline
General &
\begin{tabular}[c]{@{}l@{}}
3D TLC NAND flash-based SSD; \\
4 ARM Cortex-A9 cores; NVMe 2.0
\end{tabular} \\ \hline
I/O & PCIe 4.0 $\times$4 (8 GB/s) \\ \hline
DRAM & 4 GB LPDDR5, 51.2 GB/s \\ \hline
Flash Memory &
\begin{tabular}[c]{@{}l@{}}
16 Channels/SSD; 4 Chips/Channel; 2.4 GB/s
\end{tabular} \\ \hline
Flash Parameters &
\begin{tabular}[c]{@{}l@{}}
Page size = 32 KB; 2 Dies/Chip; 2 Planes/Die; \\ 1024 Blocks/Plane; 1024 Pages/Block; 8-bit Channel Bus
\end{tabular} \\ \hline
\end{tabular}
}
\label{tab:sim_config}
\end{table}

\begin{figure*}[t]
    \centering
    \epsfig{file=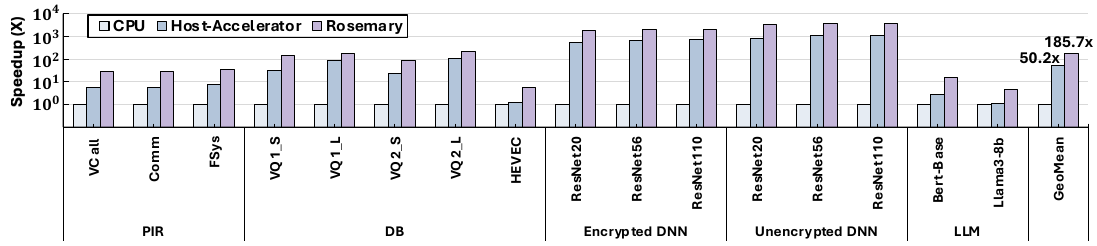, width=\textwidth}
    \caption{Overall speedup.}
    \label{fig:overall_ssd}
\end{figure*}

\subsection{Baselines}
\label{sec:baselines}
\textbf{CPU:} We evaluate open-source implementations on an AMD Ryzen Threadripper 9970X with 256GB RAM, using swap files when datasets exceed memory capacity.

\noindent\textbf{Host-Accelerator (HA):} Our baseline couples SSD storage with a state-of-the-art FHE accelerator~\cite{ufc}, streaming data directly from SSD to accelerator HBM over PCIe~\cite{gdrcopy} and optimistically assuming full PCIe bandwidth. To validate the model, we compare the non-SSD portion of our simulation against the original paper and find less than 8\% difference.

\noindent\textbf{INSPIRE:} We use INSPIRE~\cite{inspire} as the baseline for ASIC-based FHE ISP acceleration. To ensure a faithful comparison, we validate our INSPIRE model against the original paper.

\subsection{Workloads}
\label{sec:worklods}
We evaluate \Design on diverse FHE workloads, following state-of-the-art open-source implementations and preserving their algorithms, operator structure, bootstrapping placement, and $\geq$128-bit-secure parameters~\cite{fhe_standard_22}. We port them into our framework as follows:

\textbf{\textit{(1)} Private Information Retrieval (PIR):} We evaluate PIR using the \textbf{SmartPIR} protocol~\cite{smartpir}. We map its BFV parameters~\cite{bfv}, $(N,T,\log Q)=(2^{13},2^{16}+1218)$, to BGV with $N=4096$ and $\log QP=218$, supporting multiplicative depth 2. We evaluate on VCall ($2^{32}$ entries, 96B records), Comm ($2^{30}$, 288B), and Fsys ($2^{17}$, 10MB)~\cite{inspire}. 

\textbf{\textit{(2)} Database (DB):} We build our database benchmarks on \textbf{Engorgio}~\cite{engorgio}\footnote{\url{https://github.com/Errantry73/Engorgio}.} and \textbf{HEVEC}~\cite{llm_database}\footnote{\url{https://github.com/cloneisyou/HEVEC}.}, both CKKS-based designs with customized or quantized encoding. \Design supports them through its generic FHE primitive support. For Engorgio, we evaluate VQ1 and VQ2 (Section~6.3 and Appendix~D in ~\cite{engorgio}) with 10K dataset entries. The workload includes homomorphic distance, synchronization, and non-linear top-$k$ queries with bootstrapping, using parameters $L=43$, $N=2^{17}$, and $\log QP=3516$. \textbf{HEVEC} is an HE database implementation~\cite{llm_database} with coefficient encoding (Section 4 in~\cite{llm_database}) with $N=4096$ and $\log QP=109$, evaluate it on Deep1B~\cite{deep1b}.

\textbf{\textit{(3)} DNN Inference:} This benchmark includes \textbf{ResNet}~\cite{ppml_22_2}\footnote{\url{https://github.com/snu-ccl/FHE-MP-CNN}.} on CIFAR-10. We follow the algorithmic implementation and FHE parameters in~\cite{ppml_22_2}, consistent with prior accelerators~\cite{craterlake, sharp, ufc}, and evaluate both encrypted- and plaintext-weight settings.

\textbf{\textit{(4)} LLM Inference:} \textbf{NEXUS}~\cite{llm_nexus}\footnote{\url{https://github.com/zju-abclab/NEXUS/tree/main}.} provides an FHE implementation optimized for dominant attention-layer kernels, including plaintext–ciphertext matrix multiplications, ciphertext–ciphertext matrix multiplications, and bootstrapping placement. We evaluate both BERT-Base and Llama3-8B. 

\section{Evaluation}
\subsection{Overall Performance}
\subsubsection{Speedup}
\label{sec:speedup}

Figure~\ref{fig:overall_ssd} shows the performance of \Design, CPU, and Host-Accelerator (HA) across the FHE workloads in Section~\ref{sec:worklods}. \Design outperforms the CPU/GPU baselines by $4.6\times$ to $3790.9\times$, achieving a GM speedup of $185.7\times$. Relative to HA, \Design delivers up to $5.4\times$ speedup and a GM speedup of $3.7\times$. The large spread in speedup over CPU baselines mainly arises from differences in the platforms and assumptions used in prior work. For instance, the $3790.9\times$ speedup on ResNet is partly driven by a single-threaded SEAL~\cite{seal} CPU baseline. Meanwhile, gains over recent GPU-based implementations are smaller, at $2.8\times$ for BERT-Base and $1.1\times$ for Llama3-8B, as these baselines already employ strong GPU backends and incur large plaintext-weight footprints.\footnote{The Nexus open-source LLM implementations use 4$\times$A100 GPUs.} Overall, more I/O-bound workloads benefit less from HA, which accelerates computation but not storage I/O.

Compared to HA, \Design provides larger gains on data-intensive workloads. For PIR, it achieves an average speedup of $4.3\times$ across \textsc{VCall}, \textsc{Comm}, and \textsc{FSys}. For the database workloads, \textsc{VQ1\_S} and \textsc{VQ2\_S} achieve $4.6\times$ and $3.6\times$ speedups over HA, while \textsc{VQ1\_L} and \textsc{VQ2\_L} see more moderate gains of around $2\times$. The smaller improvement for \textsc{VQ\_L} comes from its higher baseline slot usage. Since each ciphertext has a fixed number of slots, \Design can pack fewer ciphertexts together and obtain less benefit from in-SSD reduction. \textsc{VQ2} also benefits less than \textsc{VQ1} because its 16-way sort is more constrained by host-side computation than the 4-way sort in \textsc{VQ1}. Among the database workloads, \textsc{HEVEC} achieves the largest speedup over HA because its coefficient-encoded inner product relies primarily on lightweight multiply-add operations while substantially reducing data movement.

For encrypted and unencrypted DNN inference, \Design achieves speedups of $2.95\times$ to $3.52\times$ and $3.40\times$ to $4.19\times$ compared to HA, respectively. These results show that \Design is not limited to retrieval-style workloads, but also provides robust improvements for model inference pipelines with substantial arithmetic and large working sets. \Design improves over HA by $5.37\times$ on BERT-Base and $4.09\times$ on Llama3-8B. The performance gains of \Design over HA arise from two main factors: (1) in-storage execution minimizes host-side I/O by reducing ciphertext transfers across the host-storage interface; and (2) heterogeneous execution exposes additional compute parallelism between the host FHE accelerator and the RosemarySSD FHE engine.

\begin{figure}
    \raggedleft
    \epsfig{file=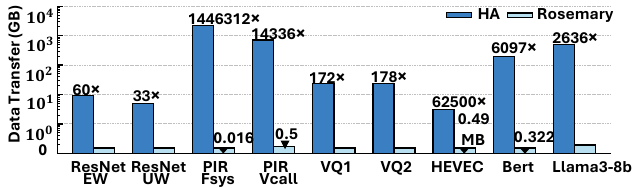, width=0.95\columnwidth}
    \caption{PCIe traffic comparison.} 
    \label{fig:pcie_traffic}
\end{figure}

\subsubsection{PCIe Traffic}
\label{sec:pcie_traffice}
Figure~\ref{fig:pcie_traffic} compares the PCIe traffic of HA and \Design. Across all workloads, \Design reduces host--SSD data movement by a geometric mean of $2207.7\times$. The largest reductions occur for data-intensive workloads such as \textsc{PIR} and \textsc{HEVEC}, where communication dominates end-to-end execution. LLMs also show substantial reductions in weight traffic, since state-of-the-art implementations often replicate weights to lower compute complexity and accommodate rotation-heavy execution. The results for \textsc{VQ1} and \textsc{VQ2} are conservative because we evaluated 10K dataset entries, which is much smaller than real-world deployment scales.

\subsection{Comparison with Previous ISP Design}
\label{sec:eval_prev}

\begin{figure}
    \epsfig{file=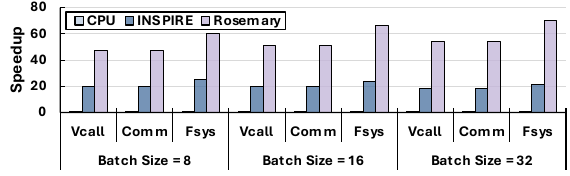, width=\columnwidth}
    \caption{Performance comparison between \Design and INSPIRE~\cite{inspire}.} 
    \label{fig:prior_pir}
\end{figure}

Figure~\ref{fig:prior_pir} compares INSPIRE~\cite{inspire} and \Design on \textsc{PIR} workload with various batch sizes. Overall, across all dataset and batch size combinations, INSPIRE achieves a GM speedup of $20.6\times$ over the CPU baseline, whereas \Design achieves a $2.7\times$ higher GM speedup than INSPIRE. Here, we re-normalize both systems under the CPU baseline. The speedups of both systems are more conservative compared to the original results in ~\cite{inspire}, as our CPU has a higher frequency and is associated with PCIe4x4 instead of PCIe3x4. We do not directly compare against SmartPIR because its reported results are obtained under a different system stack, which SmartPIR combines protocol-level and CSD-level optimizations on a commercial CSD array, including zero-skipping for variable-length entries and internal peer-to-peer communication support. \Design outperforms INSPIRE more significantly at larger batch sizes. The speedup grows from $2.4\times$ at batch size 8 to $3.1\times$ at batch size 32. The reason is that \Design sustains higher internal processing throughput and can absorb larger batches without substantial slowdown, whereas INSPIRE degrades as more queries become active because each processing engine has limited buffer capacity.


\subsection{Ablation Studies}
\label{sec:ablation}

\begin{figure}
    \epsfig{file=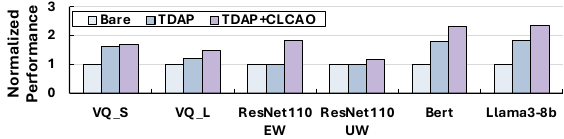, width=\columnwidth}
    \caption{The ablation study of our proposed optimizations on \Design.} 
    \label{fig:ablation}
\end{figure}

We perform an ablation study to isolate the benefits of the two main compiler optimizations in \Design: TDAP and CLCAO. Figure~\ref{fig:ablation} compares three variants. \textit{Bare} disables both optimizations, \textit{TDAP} enables only heterogeneous partitioning, and \textit{TDAP+CLCAO} enables both. Overall, enabling TDAP alone improves performance by a geometric mean of $1.37\times$ over \textit{Bare}, and the full \textit{TDAP+CLCAO} configuration improves performance by $1.75\times$ on average, corresponding to an additional $1.28\times$ gain over TDAP alone. 

TDAP benefits more for workloads with substantial cross-device communication or sophisticated execution phases, where our baseline allocation manually hints to the compiler and assigns all non-linear FHE segments to the \textit{FHE Accel} and linear segments to \textit{RosemarySSD}. For example, TDAP provides $1.62\times$ and $1.21\times$ speedup on \textsc{VQ\_S} and \textsc{VQ\_L}, and $1.81\times$ and $1.85\times$ on \textsc{BERT} and \textsc{Llama3-8B}, respectively. These gains show that cost-aware placement avoids unfavorable offloadings and unnecessary host--SSD transfers. CLCAO further improves efficiency by refining the SSD-resident segments selected by TDAP. Across the evaluated workloads, the full configuration reaches up to $2.35\times$ speedup over \textit{Bare}. The additional benefit is especially large for \textsc{ResNet110-EW}, where CLCAO increases performance from no gain under TDAP alone to $1.84\times$. Without CLCAO, the convolutional matrix multiplications and weight-input \texttt{HMult} operations repeatedly fetch new keys and evict recently used ones. \textsc{ResNet} workloads see little benefit from TDAP, as the manual strategy is already well explored and tested. In contrast, for LLM workloads, the full stack achieves $2.30\times$ speedup on \textsc{BERT} and $2.35\times$ on \textsc{Llama3-8B}.

\subsection{Architectural Analysis}
\label{sec:arch_analysis}
\subsubsection{Utilization}
\label{sec:utilization}

\begin{figure}
    \epsfig{file=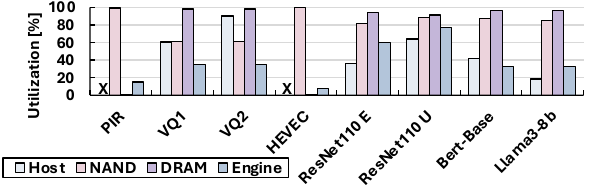, width=\columnwidth}
    \caption{Utilization of \Design components.} 
    \label{fig:utilization}
\end{figure}

Figure~\ref{fig:utilization} shows the utilization of the host, NAND, LPDDR DRAM, and FHE engine. Overall, the high utilization of NAND or DRAM bandwidth indicates that efficient data movement and buffering are critical to end-to-end performance. For compute-light workloads such as \textsc{PIR} and \textsc{HEVEC}, NAND is the primary bottleneck, reaching 99.2\% and 99.8\% utilization, while FHE-engine utilization remains low at 14.8\% and 7.6\%, reflecting their data-dominated behavior. For DB workloads, \textsc{VQ1} and \textsc{VQ2} place the highest pressure on LPDDR DRAM, both reaching 98.2\% utilization, with host utilization also high at 60.8\% and 90.2\%. Their moderate FHE-engine utilization suggests that performance is limited more by intermediate ciphertext movement than by computation. Neural-network workloads are relatively more compute-intensive. Unencrypted \textsc{ResNet110} achieves the highest FHE-engine utilization at 77.3\%, while encrypted \textsc{ResNet110} reaches 59.9\%. \textsc{BERT-Base} and \textsc{Llama3-8B} still show high NAND and DRAM utilization but moderate engine utilization, indicating that weight and key movement remain a major bottleneck.

\subsubsection{Scratchpad size}
\label{sec:edram_size}

\begin{figure}
    \epsfig{file=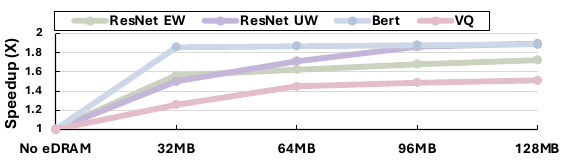, width=0.95\columnwidth}
    \caption{Sensitivity to scratchpad size in \Design.} 
    \label{fig:edram_size}
\end{figure}

Increasing eDRAM improves performance by buffering more intermediate ciphertexts. Compared to the design without eDRAM, 32MB, 64MB, 96MB, and 128MB scratchpads improve performance by \(1.53\times\), \(1.66\times\), \(1.72\times\), and \(1.75\times\) on average, respectively. The gain saturates beyond moderate capacities. For example, \textsc{BERT} already achieves \(1.86\times\) speedup at 32MB and improves only marginally to \(1.89\times\) at 128MB.

\subsubsection{\revise{FHE engine}}
\label{sec:fhe_engine_size}

\begin{figure}
    \epsfig{file=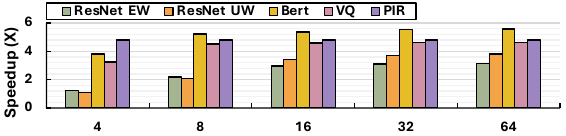, width=0.95\columnwidth}
    \caption{\revise{Sensitivity to FHE engine size in \Design.}} 
    \label{fig:fhe_engine_size}
\end{figure}

\revise{We further evaluate \Design's sensitivity to ISP FHE-engine size by varying the number of arithmetic units (AUs). With 4, 8, 16, 32, and 64 AUs, \Design achieves geometric-mean speedups of $2.4\times$, $3.5\times$, $4.1\times$, $4.3\times$, and $4.3\times$, respectively, over the \textbf{HA} baseline. Performance scales substantially up to 16 AUs, but the gains diminish beyond 32 AUs as execution becomes increasingly constrained by the available NAND bandwidth and other components.}

\subsubsection{\revise{Scalability}}
\label{sec:scalability}

\begin{figure}
    \epsfig{file=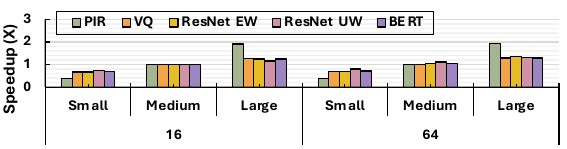, width=0.95\columnwidth}
    \caption{\revise{\Design scalability with 16 AUs. The results are normalized to the default 16-AUs with the Medium SSD.}} 
    \label{fig:scalability}
\end{figure}

\revise{Figures~\ref{fig:scalability} evaluate \Design scalability with 16- and 64-AU engines. The Small, Medium, and Large SSDs provide 8, 16, and 32 channels at 1.8, 2.4, and 2.4,GB/s per channel, respectively, with the Large configuration reaching 76.8,GB/s aggregate bandwidth and using LPDDR6 DRAM. Smaller SSDs substantially degrade performance for both engine sizes, especially for PIR, which nearly saturates NAND bandwidth. The Large SSD improves geometric-mean performance by about 38\%.}

\subsection{Area and Power Overhead}
\label{sec:area_power}
\subsubsection{Area Overhead}
Table~\ref{tab:area_power} summarizes the area and power breakdown of \Design. At 7nm, \Design occupies 45.6mm\textsuperscript{2} and consumes 16.7W, only 2.3\% of the total area of a Samsung 990 Pro~\cite{990pro4tb}. \revise{This is comparable in scale to prior ISP designs~\cite{ndsearch, beacongnn}, with Beacongnn employs 27.7mm\textsuperscript{2} and NDSEARCH employs 43.09mm\textsuperscript{2} total area of the customized logic. The area and power employment is much smaller than SHARP~\cite{sharp} (149.2mm\textsuperscript{2}/66.2W), excluding their HBM PHY overhead.} SmartSSD~\cite{smartssd}, by comparison, requires about 800mm\textsuperscript{2} of logic area alone. NTTU and MODU account for 21.1\% of the area and 44.9\% of power. SRAM and scratchpad contribute 16.2\%/19.8\% and 38.2\%/11.4\% of total area/power, respectively. The interconnect occupies a notable area to support generic NTT data movement, as AUs need to support for complex NTT data patterns across generic parameter sets to maintain high utilization for different schemes. \revise{The mProgram dispatcher, descriptor lookup, command queues, and arbitration logic together occupy less than 0.1mm\textsuperscript{2}.}

\begin{table}[t]
\centering
\caption{Area and power breakdown of \Design.}
\label{tab:area_power}
\setlength{\tabcolsep}{4pt}
\renewcommand{\arraystretch}{0.95}
\begin{tabular}{@{}lrr@{}}
\toprule
\textbf{Component} & \textbf{Area (mm\textsuperscript{2})} & \textbf{Power (W)} \\
\midrule
NTTU & 6.5   & 4.8 \\
MODU & 3.3   & 2.7 \\
OT \& KeyGen & 0.075 & 0.063 \\
SRAM & 7.4   & 3.3 \\
Scratchpad & 17.4  & 1.9 \\
Interconnect \& Others & 10.9 & 3.9 \\
\midrule
\textbf{Total} & \textbf{45.6} & \textbf{16.7} \\
\bottomrule
\end{tabular}
\end{table}

\subsubsection{Power Budget and Energy Efficiency} \revise{ISP accelerators} are bounded by the PCIe power budget, about 55W~\cite{beacongnn, ndsearch}. Our total power consumption of 16.7W stays within this budget. \revise{Recent server-oriented EDSFF E3.S devices support power envelopes up to 25,W for standard form factors and 40,W for thicker variants, including computational-storage and accelerator devices~\cite{kioxia_edsff_e3}. Their enclosure and heatsink designs are optimized for efficient heat transfer to server airflow~\cite{kioxia_edsff_e3, samsung_pm1743}. At 16.7,W, \Design remains below the standard 25,W E3.S envelope, leaving 8.3,W of headroom and fitting within existing server cooling capabilities.} 

\begin{figure}
    \epsfig{file=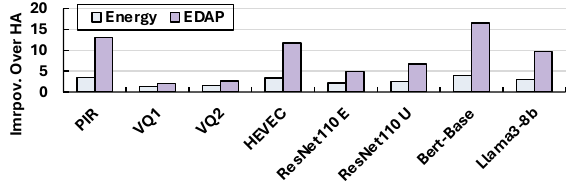, width=0.95\columnwidth}
    \caption{Energy and cost efficiency of \Design.} 
    \label{fig:edap}
\end{figure}

As shown in Figure~\ref{fig:edap}, \Design improves energy efficiency by up to 3.9$\times$. The energy efficiency improvement is mainly from the decrease in PCIe data movement and overall latency reduction. Relative to HA, \Design eliminates substantial SSD I/O by reducing feature-vector transfers. We also evaluate the cost of \Design Energy-Delay-Area-Product (EDAP). \Design reduces EDAP ranging from 6.71$\times$ to 16.48$\times$. Overall, \Design demonstrates substantial advantages in both cost and energy efficiency, making it a practical and sustainable solution for large-scale FHE deployments.

\subsection{\revise{Case Study: Private ML Inference}}
\label{sec:case_study}

\begin{figure}
    \epsfig{file=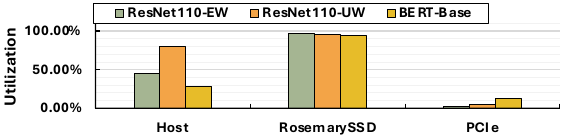, width=0.95\columnwidth}
    \caption{\revise{Overall resource utilization for representative ML inference workloads of \Design.}} 
    \label{fig:case_study}
\end{figure}

\revise{
Figure~\ref{fig:case_study} shows host-accelerator, \Design SSD, and PCIe utilization during end-to-end private ML inference under TDAP. For ResNet, TDAP keeps compute-intensive nonlinear activations on the host while offloading data-intensive linear layers to ISP. The host is active for 45.0\% and 80.2\% of execution in ResNet110-EW and ResNet110-UW. The higher utilization for UW arises because plaintext-weight multiplications avoid key switching. For BERT-Base, TDAP executes LayerNorm on the host and offloads large attention and feed-forward linear layers to ISP, where their large weight and ciphertext footprints make near-storage execution advantageous.
}

\section{Related Work}
\label{sec:related_work}

\noindent\textbf{FHE Acceleration:} Prior FPGA, GPU, and ASIC accelerators substantially improve FHE latency and throughput~\cite{cheddar, neo, multigpu, fideslib, craterlake, sharp, ufc, cinnamon}, while compiler frameworks optimize program generation and bootstrapping placement~\cite{fhelipe, dacapo}. These efforts primarily target computation, whereas \Design addresses the storage I/O bottleneck of large-scale, data-intensive FHE workloads.

\noindent\textbf{Near-Data Processing for HE:} Memory-centric accelerators reduce main-memory traffic but remain capacity-limited~\cite{fhendi, anaheim}. In/near-storage designs reduce I/O for specialized workloads such as PIR, PSI, and string matching~\cite{inspire, smartpir, ares, ciphermatch}, while SmartSSD-style platforms are constrained by limited on-device compute~\cite{poseidon_ndp, smartssd, csd1}. \Design instead combines ISP with host-side heterogeneous compute to support general FHE workloads with reduced data movement.

\section{Conclusion}
\label{sec:conclusion}
This work presents \Design, an efficient and adaptable in-storage FHE acceleration system for large-scale applications. \Design combines a novel ISP architecture, \Design SSD, with hardware--software co-design through Compiler-Level Ciphertext-Aware Optimizer (CLCAO) and Throughput-Data-Aware Program Allocator (TDAP). Together, they enable efficient offloading of data-intensive FHE stages and reduce PCIe data movement. \Design achieves gmean speedups of $185.7\times$ and $3.7\times$ over state-of-the-art CPU/GPU implementations and the UFC~\cite{ufc} baseline, respectively, while improving energy efficiency by up to $3.9\times$ and EDAP by $16.5\times$.

\section{ACKNOWLEDGEMENTS}
The authors used the generative AI tool (ChatGPT) solely to polish and improve the readability of the initial draft. All technical content, experimental results, and interpretations were produced and verified by the authors.


\bibliographystyle{IEEEtranS}
\bibliography{refs}

\vskip -2\baselineskip plus -1fil
\begin{IEEEbiography}[{\includegraphics[width=1in,height=1.25in,clip,keepaspectratio]{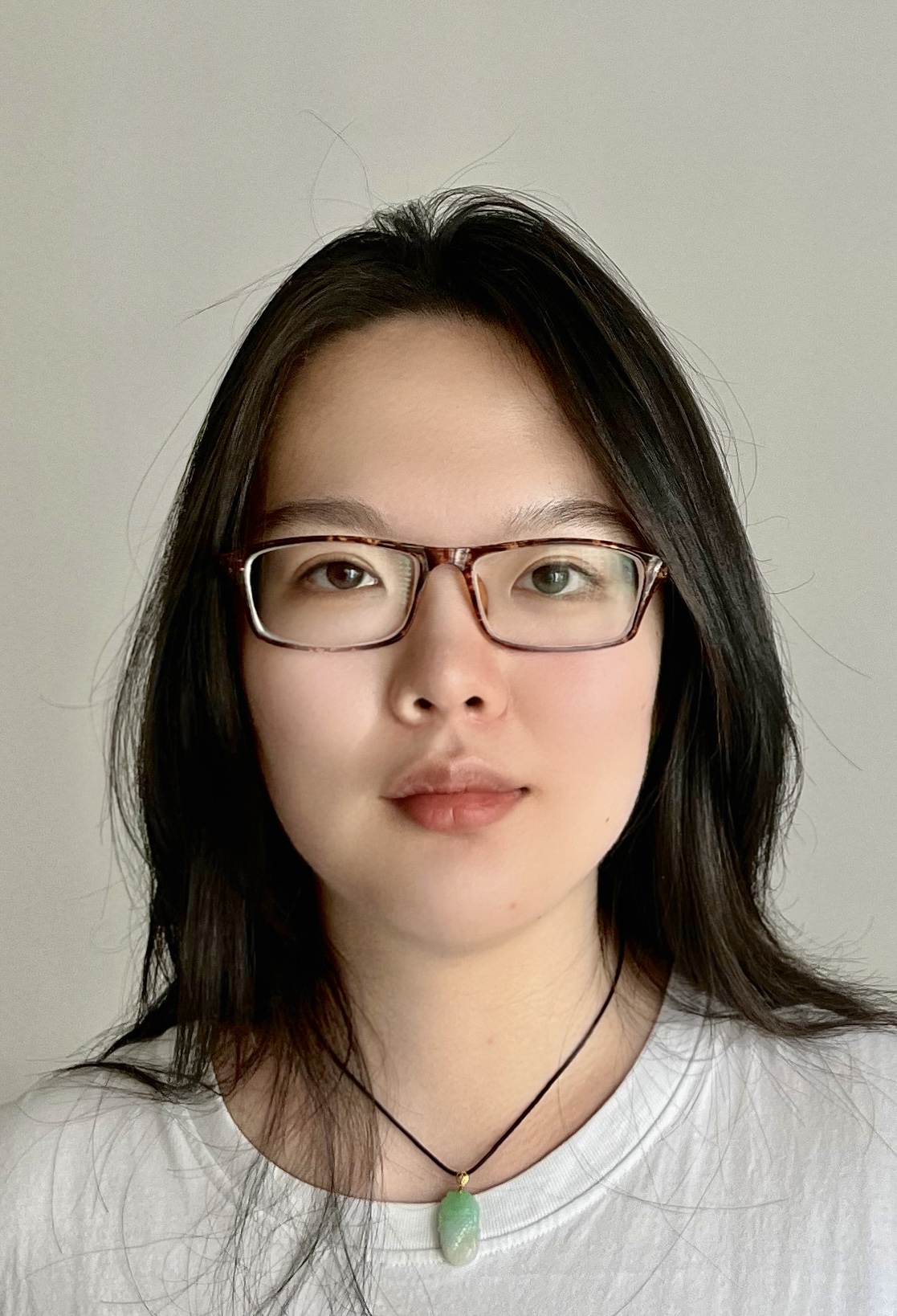}}]{Xuan Wang} received her B.S. degree from the University of California at San Diego, La Jolla, CA, in 2023. She has been pursuing the Ph.D. degree in Computer Science and Engineering at the University of California at San Diego, La Jolla, CA, since 2023. Her research interests include hardware-software co-design, computer architecture, fully homomorphic encryption, and privacy-preserving computing.
\end{IEEEbiography}

\vskip -2\baselineskip plus -1fil
\begin{IEEEbiography}[{\includegraphics[width=1in,height=1.25in,clip,keepaspectratio]{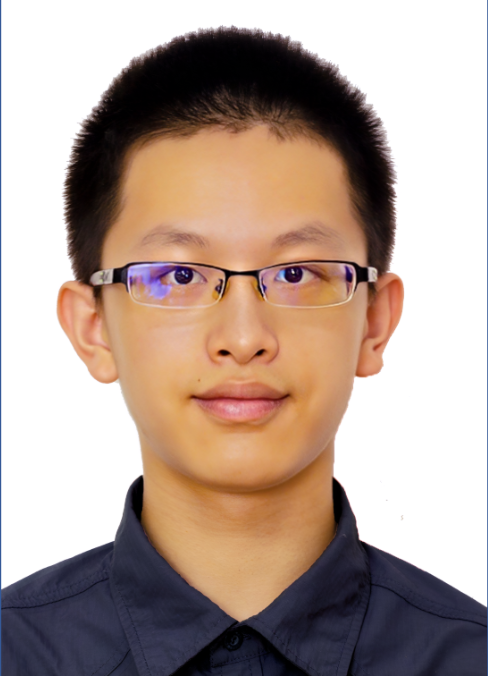}}]{Tianqi Zhang} received the B.S. degree from the University of Electronic Science and Technology of China, Sichuan, China. He is currently pursuing the Ph.D. degree in computer science and engineering with the University of California at San Diego, La Jolla, CA, USA, since 2023. His research interests include algorithm and hardware co-design, neuromorphic computing, and domain-specific accelerators.
\end{IEEEbiography}

\vskip -2\baselineskip plus -1fil
\begin{IEEEbiography}[{\includegraphics[width=1in,height=1.25in,clip,keepaspectratio]{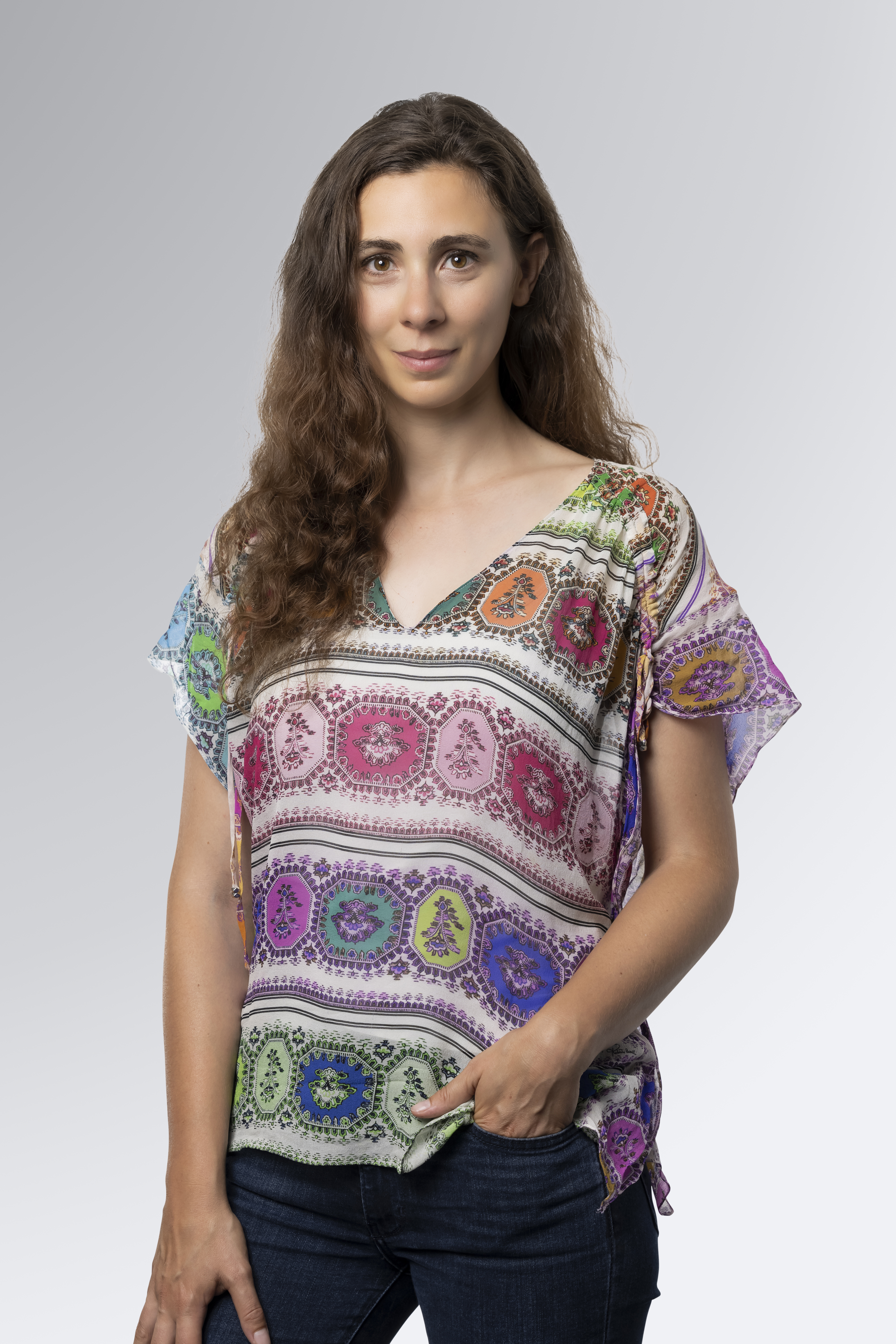}}]{Gabrielle De Micheli} is a Staff Cryptography Research Scientist at LG Electronics, North America, conducting research focusing on fully homomorphic encryption and more generally post-quantum cryptography. She also has an affiliated research appointment as Visiting Scholar at UC San Diego in the Department of Electrical and Computer Engineering. She completed her PhD at INRIA, France, in 2021, and was a postdoctoral scholar at UC San Diego from 2021-2024. 
\end{IEEEbiography}

\vskip -2\baselineskip plus -1fil
\begin{IEEEbiography}[{\includegraphics[width=1in,height=1.25in,clip,keepaspectratio]{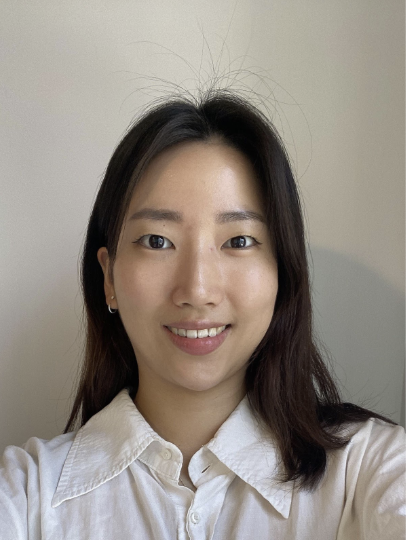}}]{Yujin} received the BS degree in electrical and computer engineering from Seoul National University in 2020. She is currently working toward the PhD degree in computer science and engineering with the University of California, San Diego. Her research focuses on system security and data privacy. Especially, she works on privacy-preserving machine learning (PPML) with applications of cryptographic technologies like fully homomorphic encryption.
\end{IEEEbiography}

\vskip -2\baselineskip plus -1fil
\begin{IEEEbiography}[{\includegraphics[width=1in,height=1.25in,clip,keepaspectratio]{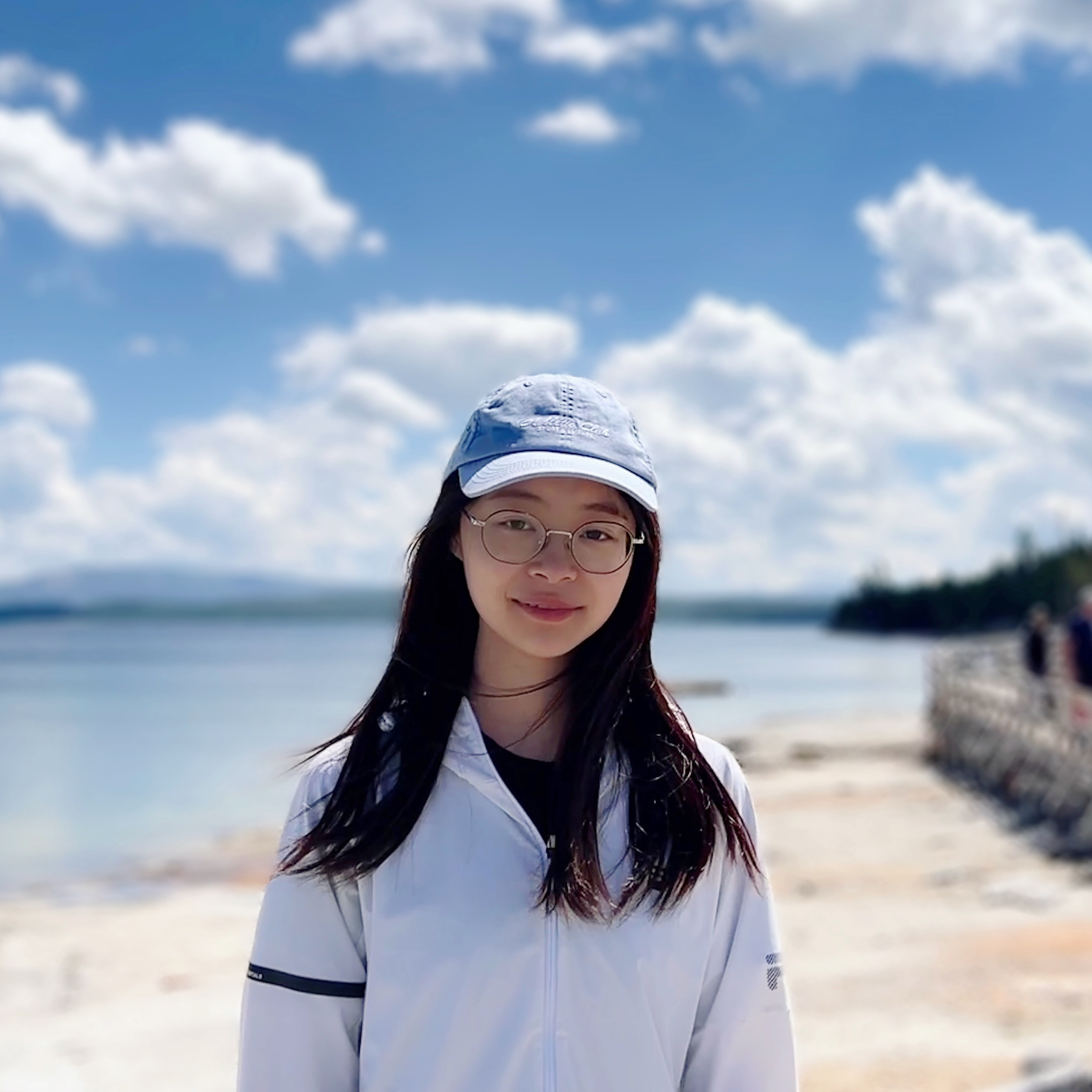}}]{Keming Fan} received the B.E. degree from ShanghaiTech University, Shanghai, China. She is currently pursuing the Ph.D. degree in electrical and computer engineering with the University of California at San Diego, La Jolla, CA, USA, in 2022. Her research interests include hardware-software co-design, mixed-signal circuit design, and in-memory computing.
\end{IEEEbiography}

\vskip -2\baselineskip plus -1fil
\begin{IEEEbiography}[{\includegraphics[width=1in,height=1.25in,clip,keepaspectratio]{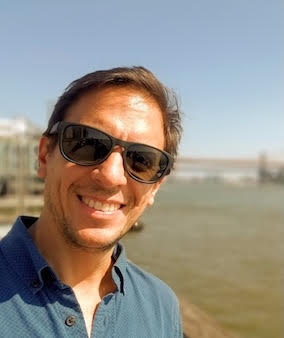}}]{Augusto Vega} is a Senior Research Scientist in AI hardware and computer architecture. His work centers on the co-design of machine learning systems, accelerators, and scalable computing platforms, with an emphasis on performance, efficiency, and deployment at scale. His research interests include AI infrastructure, workload-driven architecture design, system optimization, and emerging technologies for next-generation intelligent computing systems. He holds a Ph.D. degree from Polytechnic University of Catalonia (UPC), Spain.
\end{IEEEbiography}

\vskip -2\baselineskip plus -1fil
\begin{IEEEbiography}[{\includegraphics[width=1in,height=1.25in,clip,keepaspectratio]{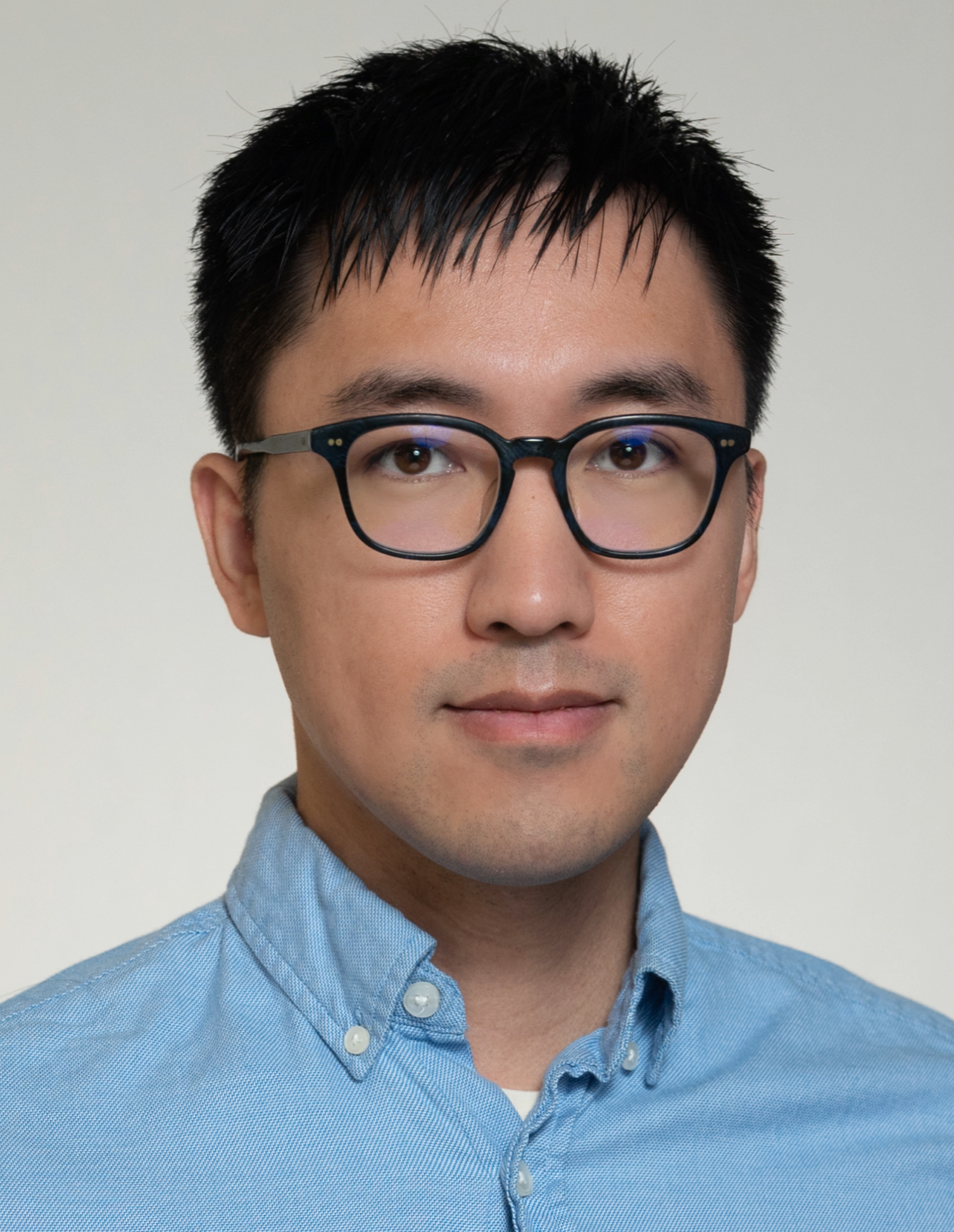}}]{Minxuan Zhou}
(Member, IEEE) received his Ph.D. degree from University of California San Diego, CA, USA, in 2023. He is an Assistant Professor of Computing at Illinois Institute of Technology, Chicago, IL, USA. His research interests include computer architecture, software-hardware co-design, machine learning acceleration, and privacy-preserving computing.
\end{IEEEbiography}

\vskip -2\baselineskip plus -1fil
\begin{IEEEbiography}[{\includegraphics[width=1in,height=1.25in,clip,keepaspectratio]{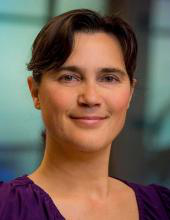}}]{Tajana Rosing}
(Fellow, IEEE) received her Ph.D. degree from Stanford University, Stanford, CA, USA, in 2001. She is a Professor, a Holder of the Fratamico Endowed Chair, and the Director of System Energy Efficiency Laboratory, University of California at San Diego, La Jolla, CA, USA. She is leading a number of projects, including efforts funded by DARPA/SRC JUMP 2.0 PRISM program with focus on design of accelerators for analysis of big data, DARPA and NSF funded projects on hyperdimensional computing, and SRC funded project on IoT system reliability and maintainability. Her current research interests include energy-efficient computing, cyber–physical, and distributed systems.
\end{IEEEbiography}

\end{document}